\documentclass[a4paper,11pt]{article}
\pdfoutput=1

\usepackage[utf8]{inputenc}
\usepackage{amsmath,amssymb,mathtools}
\usepackage{graphicx}
\usepackage{jheppub}
\usepackage{xspace}

\allowdisplaybreaks

\DeclareRobustCommand{\ensuremathrm}[1]{\ensuremath{\mathrm{#1}}\xspace}

\DeclareRobustCommand{\ensuremathcal}[1]{\ensuremath{\mathcal{#1}}\xspace}
\DeclareRobustCommand{\rd}{\ensuremathrm{d}} 
\DeclareRobustCommand{\ord}{\ensuremathcal{O}\xspace} 
\DeclareRobustCommand{\obs}{\ensuremathcal{O}\xspace} 

\DeclareRobustCommand{\rs}{\ensuremathrm{s}}
\DeclareRobustCommand{\rT}{\ensuremathrm{T}}
\DeclareRobustCommand{\jet}{\ensuremathrm{jet}\xspace}
\DeclareRobustCommand{\cut}{\ensuremathrm{cut}\xspace}
\DeclareRobustCommand{\alphas}{\ensuremath{\alpha_\rs}\xspace}
\DeclareRobustCommand{\mur}{\ensuremath{\mu_\mathrm{R}}\xspace}
\DeclareRobustCommand{\muR}{\mur}
\DeclareRobustCommand{\muf}{\ensuremath{\mu_\mathrm{F}}\xspace}
\DeclareRobustCommand{\muF}{\muf}

\DeclareRobustCommand{\PH}{{\ensuremathrm{H}}\xspace}
\DeclareRobustCommand{\PZ}{{\ensuremathrm{Z}}\xspace}
\DeclareRobustCommand{\PW}{{\ensuremathrm{W}}\xspace}

\DeclareRobustCommand{\Pgg}{{\ensuremathrm{\gamma}}\xspace}

\DeclareRobustCommand{\Pe}{{\ensuremathrm{e}}\xspace}
\DeclareRobustCommand{\Pep}{{\ensuremathrm{e^+}}\xspace}
\DeclareRobustCommand{\Pem}{{\ensuremathrm{e^-}}\xspace}

\DeclareRobustCommand{\Pqt}{{\ensuremathrm{t}}\xspace}
\DeclareRobustCommand{\Paqt}{{\ensuremathrm{\bar{t}}}\xspace}
\DeclareRobustCommand{\Pp}{{\ensuremathrm{p}}\xspace}

\DeclareRobustCommand{\GeV}{\ensuremathrm{GeV}\xspace}

\DeclareRobustCommand{\NNLOJET}{\textsc{NNLOjet}\xspace}
\DeclareRobustCommand{\ZEUS}{\text{ZEUS}\xspace}
\DeclareRobustCommand{\LO}{\text{LO}\xspace}
\DeclareRobustCommand{\NLO}{\text{NLO}\xspace}
\DeclareRobustCommand{\NNLO}{\text{NNLO}\xspace}
\DeclareRobustCommand{\NNNLO}{\text{N${}^{3}$LO}\xspace}
\DeclareRobustCommand{\PtoB}{\text{P2B}\xspace}
\DeclareRobustCommand{\Next}{\text{N}\xspace}
\DeclareRobustCommand{\Born}{\text{B}\xspace}
\DeclareRobustCommand{\incl}{\text{incl.}\xspace}
\DeclareRobustCommand{\intPhi}[1]{\ensuremath{\int_{\mathrlap{\Phi_{#1}}}\quad}\xspace}

\title{\boldmath\texorpdfstring{\NNNLO}{N3LO} Corrections to Jet Production in Deep Inelastic Scattering using the Projection-to-Born Method}
\author[a]{J.\ Currie,}
\author[b]{T.\ Gehrmann,}
\author[a]{E.W.N.\ Glover,}
\author[c]{A.\ Huss,}
\author[a]{J.\ Niehues,}
\author[d]{A.\ Vogt}
\affiliation[a]{Institute for Particle Physics Phenomenology, Durham University, Durham, DH1 3LE, UK}
\affiliation[b]{Department of Physics, Universität Zürich, Winterthurerstrasse 190, 8057 Zürich, Switzerland}
\affiliation[c]{Theoretical Physics Department, CERN, 1211 Geneva 23, Switzerland}
\affiliation[d]{Department of Mathematical Sciences, University of Liverpool, Liverpool L69 3BX, UK}
\emailAdd{james.currie@durham.ac.uk}
\emailAdd{thomas.gehrmann@uzh.ch}
\emailAdd{e.w.n.glover@durham.ac.uk}
\emailAdd{alexander.huss@cern.ch}
\emailAdd{jan.m.niehues@durham.ac.uk}
\emailAdd{andreas.vogt@liverpool.ac.uk}
\abstract{%
  Computations of higher-order QCD corrections for processes with exclusive final states require a subtraction method for real-radiation contributions.
  We present the first-ever generalisation of a subtraction method for third-order (\NNNLO) QCD corrections. 
  The Projection-to-Born method is used to combine inclusive \NNNLO coefficient functions with an exclusive second-order (NNLO) calculation for a final state with an extra jet.
  The input requirements, advantages, and potential applications of the method are discussed, and validations at lower orders are performed.
  As a test case, we compute the \NNNLO corrections to kinematical distributions and production rates for 
  single-jet production in deep inelastic scattering in the laboratory frame, and compare them with data from the ZEUS experiment at HERA. The corrections are small in the 
  central rapidity region, where they 
  stabilize the
  predictions to sub per-cent level. The corrections  increase substantially towards forward rapidity where 
  large logarithmic effects are expected, thereby yielding an improved 
  description of the data in this region.
}
\preprint{{\raggedleft%
  CERN-TH-2018-056 \\
  IPPP/18/21 \\
  ZU-TH 12/18 \\
  LTH 1155 \\
}}

\begin{document}

\setlength{\parskip}{0.15cm}

\maketitle

\section{Introduction}
\label{sec:intro}

Collider experiments have provided a wealth of precision measurements of basic low-multiplicity production processes and scattering reactions in particle physics.
An equally high level of accuracy in the theoretical predictions is required to turn the experimental data into highly accurate determinations of fundamental parameters (e.g., coupling constants and particle masses) or to use them in indirect searches for new-physics effects.
The necessary theoretical precision can be obtained by computing the relevant scattering cross sections to a sufficiently high order
in perturbation theory.

In this context, one distinguishes predictions for exclusive (sometimes also called fully differential or fiducial, depending on the specific application) cross sections and inclusive cross sections. 
Fiducial cross sections take account of the kinematical coverage and final-state reconstruction procedure of the experimental measurement, and can be compared directly with data. 
Inclusive cross sections are the result of an extrapolation to full kinematical coverage. 
This extrapolation is usually performed as part of the experimental analysis; it does however require detailed modelling and theory input, thereby introducing additional sources of uncertainty. 
Wherever possible, experimental measurements of precision observables at the LHC have started to shift their focus towards fiducial cross sections. 

The computation of higher-order corrections differs in important technical details between inclusive and exclusive cross sections. 
In inclusive cross sections, kinematical information on individual final-state products can be fully integrated over, thereby considerably reducing the number of independent scales in the problem under consideration.
Results for higher-order corrections for inclusive cross sections can be cast in the form of coefficient functions, which can often be obtained in closed analytical form. 
In contrast, predictions for exclusive cross sections need to keep track of the full final-state information in all subprocesses relevant at a given order. 
This is usually realised in the form of a parton-level event generator, which applies the experimental event reconstruction and kinematical cuts (collectively called measurement function) to all subprocesses in order to reconstruct the fully exclusive differential cross section. 
Owing to these important technical differences, inclusive cross sections can often be computed to a higher perturbative order than exclusive cross sections. 

QCD corrections for inclusive cross sections are available to the fourth order ($\Next^{4}\LO$) for $\Pep\Pem\to~\text{hadrons}$~\cite{kuhn}, and to the third order (\NNNLO) for deep inelastic scattering~\cite{mvv} and Higgs-boson production (integrated over rapidity) in gluon fusion at hadron colliders~\cite{hn3lo}. 
Building upon the results for deep inelastic scattering (DIS), corrections to this order have also been inferred for Higgs boson production in vector boson fusion~\cite{karlberg}.
In processes with hadrons in the initial state, these results for the partonic coefficient functions in principle require parton distributions accurate to \NNNLO, which will be enabled by the ongoing progress in the calculation of the Altarelli--Parisi splitting functions to this order~\cite{apn3lo}.

For fully differential exclusive cross sections, QCD corrections to the second order (next-to-next-to-leading order, NNLO) were computed for the $2\!\to\! 1$ processes vector-boson production~\cite{dynnlo,babisdy} and Higgs-boson production~\cite{babishiggs,hnnlo} already about a decade ago. 
In recent years, NNLO calculations have become available for many $2\to 2$ reactions at hadron colliders: 
$\Pp\Pp\to\Pgg\Pgg$~\cite{twogamma}, 
$\Pp\Pp\to V\PH$~\cite{vh}, 
$\Pp\Pp\to V\Pgg$~\cite{vgamma}, 
$\Pp\Pp\to \Pqt\Paqt$~\cite{czakon1,czakon2}, 
$\Pp\Pp\to \PH+j$~\cite{hjet,ourhj},
$\Pp\Pp\to \PW+j$~\cite{wjet,ourwj}, 
$\Pp\Pp\to \PZ+j$~\cite{ourzj,zjet}, 
$\Pp\Pp\to \Pgg+X$~\cite{mcfmgam},
$\Pp\Pp\to \PZ\PZ$~\cite{zz}, 
$\Pp\Pp\to \PW\PW$~\cite{ww}, 
$\Pp\Pp\to \PZ\PW$~\cite{zw} and 
$\Pp\Pp\to 2j$~\cite{2jnew}, as well as for the electron--positron collisions 
$\Pep\Pem \to 3j$~\cite{our3j,weinzierl3j} and lepton--proton processes 
$\Pe\Pp\to 1j$~\cite{abelof} 
$\Pe\Pp\to 2j$~\cite{disprl} and for the related $2\to 3$ hadron-collider process of Higgs production in vector boson fusion~\cite{vbfnnlo,vbfnnlo_ant}. 
These NNLO calculations of fully differential exclusive cross sections were enabled by substantial methodological developments~\cite{vbfnnlo,secdec,ourant,currie,stripper,trocsanyi,qtsub,njettiness} of infrared subtraction methods for the handling of singular contributions that appear in all parton-level subprocesses.

Infrared singular contributions appear in two forms: 
either as explicit poles from virtual loop corrections or in the form of implicit poles from real-radiation corrections, which turn into explicit poles only after integration over the phase space associated to the real radiation. 
An infrared subtraction method extracts the infrared poles from singular real radiation contributions, optimally in a process-independent manner, and allows them to be cancelled against the virtual corrections. 
The currently available methods require a varying level of preparation and partly allow the re-use of results from lower-order calculations in order to obtain NNLO corrections to exclusive cross sections. 

In this respect, the Projection-to-Born (\PtoB) method~\cite{vbfnnlo} is very efficient in re-purposing already available calculations: 
to compute the NNLO corrections to fully differential exclusive cross sections related to a final state $X$, it combines the next-to-leading order (NLO) calculation for differential cross sections for $X+j$ final states with the NNLO corrections to the inclusive cross section for final state $X$. 
In its practical implementation, it requires an extension of the NLO $X+j$ calculation, in which the kinematics of each $X+j$ final state is projected to an equivalent Born kinematics to construct a subtraction term that is subsequently used to regulate the singular behaviour present when the jet becomes unresolved. 
This method is applicable to all processes where NNLO corrections to the fully inclusive cross section are known differentially in the Born-level kinematics, i.e., the production of a vector boson~\cite{dyincl} or a Higgs boson, the latter both in gluon fusion~\cite{hincl} and in vector boson fusion~\cite{vbfincl}, at hadron colliders as well as deep inelastic lepton--hadron scattering~\cite{zijlstra}. 
Up to now, it has been applied in the calculation of NNLO corrections to Higgs-plus-two-jet production in vector boson fusion~\cite{vbfnnlo}. 

There has been substantial recent progress in calculating \NNNLO corrections to inclusive cross sections~\cite{hn3lo,karlberg}, and NNLO calculations are now becoming available for fully differential cross sections involving jet final states. 
By combining these two recent developments, we are now able to extend the \PtoB method to compute \NNNLO corrections to fully differential cross sections. 
This is the main purpose of this paper which is organized as follows.
Section~\ref{sec:method} summarises the different parton-level contributions to cross sections up to \NNNLO in QCD, and describes how the \PtoB method is applied to cancel their infrared singularities. 
As a proof-of-principle application of the method, we compute the \NNNLO corrections to single-jet production in the laboratory frame of deep inelastic scattering in Section~\ref{sec:dis} and compare our predictions to data from the HERA electron--proton collider in Section~\ref{sec:results}. 
Finally, Section~\ref{sec:conc} contains our conclusions.

\section{Method}
\label{sec:method}

Infrared singularities from real-radiation contributions at higher orders in perturbation theory arise from phase-space regions where one or more of the final-state particles become soft and/or collinear. 
These singularities appear only upon integration over the final-state phase space, and can be extracted from the real-radiation contributions by using an infrared subtraction method. 
The currently available methods at NNLO can be broadly classified in two types, depending on whether the divergent phase-space integral is regulated by applying cuts to prevent it to encompass the singular regions (cut-based or slicing:~\cite{qtsub,njettiness}), or whether it is regulated by introducing counter-terms that render the integrand finite in the singular regions  (counter-term-based or subtraction:~\cite{vbfnnlo,secdec,ourant,currie,stripper,trocsanyi}). 
In the latter methods, the construction of counter-terms typically exploits the known factorisation properties of the phase space and the QCD matrix elements in all singular regions.
 
The \PtoB method~\cite{vbfnnlo} is the simplest possible incarnation of a counter-term-based method.
The counter-term is given by the full matrix element itself, which is evaluated at its original phase-space point. 
The only difference between the real-radiation contribution and its counter-term is in the measurement function: 
the real-radiation contribution uses the actual measurement function that defines the exclusive cross section, while the measurement function of the counter-term is unity everywhere in phase space (i.e., projected to the Born-level kinematics of the leading-order process), corresponding to a fully inclusive cross section. 
This method can be applied provided that the complete kinematics  of the inclusive cross section for a final state $X$ can be inferred from the momenta of non-QCD particles (i.e., not requiring a recombination or clustering of momenta); this restricts its application in principle to the production of one or more colourless particles at hadron colliders and to DIS processes. 

For processes at hadron--hadron colliders, it should be kept in mind that the kinematics of the inclusive process is described by two variables:
the mass of the final state $X$ and its rapidity, and that the inclusive cross section is thus not to be confused with the total cross section (which is integrated over rapidity). 
Conceptually, the \PtoB method can be extended for these processes to any order in perturbation theory, provided the ingredients (the inclusive cross section at the desired order, and the fully differential cross section with an extra jet at one order below) are available:
\begin{equation}
  \frac{\rd\sigma_X^{\Next^k\LO}}{\rd\obs} \;=\; 
    \frac{\rd\sigma_{X+j}^{\Next^{k-1}\LO}}{\rd\obs}
  - \frac{\rd\sigma_{X+j}^{\Next^{k-1}\LO}}{\rd\obs_\Born}
  + \frac{\rd\sigma_X^{\Next^{k}\LO,\,\incl}}{\rd\obs_\Born} \; .
\end{equation}
Here $\rd\obs$ abbreviates the kinematical definition (usually multiply differential) of an infrared-safe observable, defined using the actual event-kinematics, while $\rd\obs_\Born$ is the limiting value of $\rd\obs$ if evaluated for the Born-level production process of $X$. 
The essence of the \PtoB method is to define a kinematical mapping that uniquely assigns $\obs_\Born$ to each $\obs$,
\begin{equation}
  \rd\obs \xrightarrow[\PtoB]{} \rd\obs_\Born \; .
\end{equation}

It should however be noted that the \PtoB method only accounts for the infrared subtraction of the most singular parts, i.e.\ for those contributions that turn from implicit to explicit poles when integrating out the last remaining parton. 
All other infrared cancellations have taken place already within the construction of the exclusive cross section with an extra jet at the previous order, $\rd\sigma_{X+j}^{\Next^{k-1}\LO}/{\rd\obs}$.
In this part of the calculation, a different infrared subtraction method must be applied, since the kinematics of the process with an extra jet typically does not allow one to define a fully inclusive cross section. 
In the application of the \PtoB method to the NNLO corrections to Higgs-plus-two-jet production in vector boson fusion~\cite{vbfnnlo}, the NLO corrections to Higgs-plus-three-jet production were taken from an existing calculation~\cite{vbf3j} based on the dipole subtraction method~\cite{cs}.

Fully differential NNLO corrections are now becoming available for a substantial number of jet production processes. 
These calculations can be used as ingredients to \PtoB calculations at \NNNLO accuracy, provided the corresponding inclusive cross sections are known to this order. 
To understand how the \PtoB method is implemented at this order, we note that the \NNNLO cross section for the production of a final state $X$ (which consists of $n$ particles at Born level) is assembled as follows:
\begin{align}
  \frac{\rd\sigma_X^{\NNNLO}}{\rd\obs}\; &= \phantom{{}+}
    \intPhi{n+3} \rd\sigma_X^{RRR} J(\obs_{n+3}) 
  + \intPhi{n+2} \rd\sigma_X^{RRV} J(\obs_{n+2}) 
  \nonumber\\&\quad
  + \intPhi{n+1} \rd\sigma_X^{RVV} J(\obs_{n+1}) 
  + \intPhi{n}   \rd\sigma_X^{VVV} J(\obs_{n}) \; ,
  \label{eq:sign3lo}
\end{align}
where $\rd\sigma_X^{ABC}$ denotes the parton-level contributions to the cross section from tree-level triple-real radiation ($RRR$), from double-real radiation at one loop ($RRV$), from single-real radiation at two loops ($RVV$) and from the three-loop  virtual corrections to the Born process ($VVV$). 
The virtual loop contributions are ultraviolet-renormalised. Processes with incoming partons also contain mass factorisation counter-terms; these are implicitly contained in the virtual loop corrections in the above formula. 
The mass factorisation counter-terms up to \NNNLO are constructed from the three-loop splitting functions~\cite{mvvsplit}.  
The function $J(\obs_{i})$ defines the observable under consideration for an $i$-particle final state, with the Born-level definition $J(\obs_\Born) \equiv J(\obs_{n})$, and $\intPhi{i}$ denotes the $i$-particle phase space integration. 
Each term in the above sum is separately infrared divergent, with infrared singularities arising from the virtual loop integrations and from the phase-space integrations over unresolved particles. 
The inclusive cross section depends only on the $n$-particle Born-level kinematics, and its \NNNLO contribution can be assembled as
\begin{align}
  \frac{\rd\sigma_X^{\NNNLO,\,\incl}}{\rd\obs_\Born}\; &= \phantom{{}+} 
    \intPhi{n+3} \rd\sigma_X^{RRR} J(\obs_\Born) 
  + \intPhi{n+2} \rd\sigma_X^{RRV} J(\obs_\Born)
  \nonumber\\&\quad
  + \intPhi{n+1} \rd\sigma_X^{RVV} J(\obs_\Born) 
  + \intPhi{n}   \rd\sigma_X^{VVV} J(\obs_\Born) \; .
  \label{eq:sign3loincl}
\end{align}

The parton-level processes $\rd \sigma_X^{RRR,\,RRV,\,RVV}$ also contribute to the NNLO corrections to the (fully differential) $X+j$ production. 
Provided that a jet $j$ is resolved in the final state, the infrared cancellations among these three contributions can be accomplished by an NNLO subtraction method. 
Using the antenna subtraction method~\cite{ourant,currie,hadant}, the NNLO cross section schematically reads~\cite{currie} 
\begin{align}
  \frac{\rd\sigma_{X+j}^{\NNLO}}{\rd\obs}\; &= \phantom{{}+} 
    \intPhi{n+3}\quad \Bigl( \rd\sigma_X^{RRR} J(\obs_{n+3})
      - \rd \sigma_{X+j}^{S,a} J(\obs_{n+2}) 
      - \rd \sigma_{X+j}^{S,b} J(\obs_{n+1}) \Bigr) 
  \nonumber\\&\quad
  + \intPhi{n+2}\quad \Bigl( \rd\sigma_X^{RRV} J(\obs_{n+2}) 
    - \rd\sigma_{X+j}^{T,a} J(\obs_{n+2})
    - \rd\sigma_{X+j}^{T,b} J(\obs_{n+1}) \Bigr) 
  \nonumber\\&\quad
  + \intPhi{n+1}\quad \Bigl( \rd\sigma_X^{RVV} J(\obs_{n+1}) 
    - \rd \sigma_{X+j}^{U} J(\obs_{n+1}) \Bigr) \; ,
 \label{eq:antsub}
\end{align} 
where $\rd\sigma_{X+j}^{S,\,T,\,U}$ denote the subtraction terms, constructed from antenna functions and reduced matrix elements for the three parton-level contributions. 
Using the factorisation properties of the multi-particle phase space into a phase space of lower multiplicity and an antenna phase space~\cite{hadant}, 
\begin{align}
  \rd\Phi_{n+3} &= \rd\Phi_{n+1} \times \rd\Phi_{A,2} \; , &  
  \rd\Phi_{\left\{\substack{n+3\\n+2}\right\}} &= \rd\Phi_{\left\{\substack{n+2\\n+1}\right\}} \times \rd\Phi_{A,1} \; ,
\end{align}
$\rd\sigma_{X+j}^{S,\,T}$ can be integrated such that all antenna subtraction terms in Eq.~\eqref{eq:antsub}
can be shown to add up to zero,
\begin{align}
  \intPhi{A,1} \rd\sigma_{X+j}^{S,a} &=
  - \rd\sigma_{X+j}^{T,a} \; ,&
  \intPhi{A,2} \rd\sigma_{X+j}^{S,b} +
  \intPhi{A,1} \rd\sigma_{X+j}^{T,b} &=
  - \rd\sigma_{X+j}^{U} \; .
  \label{eq:antcanc}
\end{align}

By construction, the integrations do not depend on the definition $J$ of the observable, and can thus be carried out for all antenna functions~\cite{ourant,gionata,ritzmann} in a process-independent and observable-independent manner. 
 
In the \NNNLO contribution to the exclusive cross section for final state $X$, the final state jet $j$ in Eq.~\eqref{eq:antsub} is no longer guaranteed to be resolved, but can become soft and/or collinear, thereby resulting in an infrared divergence upon phase-space integration. 
At this point, the \PtoB subtraction sets in:
\begin{align}
  \frac{\rd\sigma_{X}^{\NNNLO}}{\rd\obs}\; &=\;
    \frac{\rd\sigma_{X+j}^{\NNLO}}{\rd\obs} 
  - \frac{\rd\sigma_{X+j}^{\NNLO}}{\rd\obs_\Born} 
  + \frac{\rd\sigma_X^{\NNNLO,\,\incl}}{\rd\obs_\Born} 
  \nonumber\\[1mm]&=\phantom{{}+} 
    \intPhi{n+3}\quad \Bigl( \rd\sigma_X^{RRR} J(\obs_{n+3}) 
    - \rd\sigma_{X+j}^{S,a} J(\obs_{n+2}) 
    - \rd\sigma_{X+j}^{S,b} J(\obs_{n+1}) \Bigr) 
  \nonumber\\&\quad 
  - \intPhi{n+3}\quad \Bigl( \rd\sigma_X^{RRR} J(\obs_{n+3\to\Born})
    - \rd\sigma_{X+j}^{S,a} J(\obs_{n+2\to\Born})
    - \rd\sigma_{X+j}^{S,b} J(\obs_{n+1\to\Born}) \Bigr) 
  \nonumber\\&\quad 
  + \intPhi{n+2}\quad \Bigl( \rd\sigma_X^{RRV} J(\obs_{n+2}) 
    - \rd \sigma_{X+j}^{T,a} J(\obs_{n+2}) 
    - \rd \sigma_{X+j}^{T,b} J(\obs_{n+1}) \Bigr)
  \nonumber\\&\quad
  - \intPhi{n+2}\quad \Bigl( \rd\sigma_X^{RRV} J(\obs_{n+2\to\Born}) 
    - \rd\sigma_{X+j}^{T,a} J(\obs_{n+2\to\Born}) 
    - \rd\sigma_{X+j}^{T,b} J(\obs_{n+1\to\Born}) \Bigr)
  \nonumber\\&\quad
  + \intPhi{n+1}\quad \Bigl( \rd \sigma_X^{RVV} J(\obs_{n+1}) 
    - \rd\sigma_{X+j}^{U} J(\obs_{n+1}) \Bigr)
  \nonumber\\&\quad
  - \intPhi{n+1}\quad \Bigl( \rd \sigma_X^{RVV} J(\obs_{n+1\to\Born}) 
    - \rd\sigma_{X+j}^{U} J(\obs_{n+1\to\Born}) \Bigr)
  \nonumber\\[1mm]&\quad
  + \frac{\rd \sigma_X^{\NNNLO,\,\incl}}{\rd \obs_\Born} \; .
   \label{eq:p2bsub}
\end{align}
The agreement of the above equation with the original \NNNLO contribution~\eqref{eq:sign3lo} can be seen by using Eq.~\eqref{eq:antcanc} to eliminate all antenna subtraction terms, observing $\obs_\Born \equiv \obs_n$, and by expanding the last line using Eq.~\eqref{eq:sign3loincl}. 
With the \PtoB subtraction, the contribution from each phase-space multiplicity in Eq.~\eqref{eq:p2bsub} is manifestly finite and can be integrated numerically,
\begin{align}
  \frac{\rd \sigma_{X}^{\NNNLO}}{\rd\obs}\; &=\phantom{{}+} 
  \intPhi{n+3}\quad \Bigl(
    \rd\sigma_X^{RRR}     \bigl( J(\obs_{n+3}) - J(\obs_{n+3\to\Born}) \bigr) 
  \nonumber\\&\qquad\qquad\;
  - \rd\sigma_{X+j}^{S,a} \bigl( J(\obs_{n+2}) - J(\obs_{n+2\to\Born}) \bigr) 
  - \rd\sigma_{X+j}^{S,b} \bigl( J(\obs_{n+1}) - J(\obs_{n+1\to\Born}) \bigr) \Bigr)
  \nonumber\\&\quad
  + \intPhi{n+2}\quad \Bigl(
    \rd\sigma_X^{RRV}     \bigl( J(\obs_{n+2}) - J(\obs_{n+2\to\Born}) \bigr) 
  \nonumber\\&\qquad\qquad\;
  - \rd\sigma_{X+j}^{T,a} \bigl( J(\obs_{n+2}) - J(\obs_{n+2\to\Born}) \bigr) 
  - \rd\sigma_{X+j}^{T,b} \bigl( J(\obs_{n+1}) - J(\obs_{n+1\to\Born}) \bigr) \Bigr)
  \nonumber\\&\quad
  + \intPhi{n+1}\quad \Bigl(
    \rd\sigma_X^{RVV}   \bigl( J(\obs_{n+1}) - J(\obs_{n+1\to\Born}) \bigr)
  - \rd\sigma_{X+j}^{U} \bigl( J(\obs_{n+1}) - J(\obs_{n+1\to\Born}) \bigr) \Bigr)
  \nonumber\\[1mm]&\quad
  + \frac{\rd\sigma_X^{\NNNLO,\,\incl}}{\rd\obs_\Born} \; .
   \label{eq:p2bsubnum}
\end{align}

For completeness, we also summarise the structure of the NLO and NNLO corrections~\cite{vbfnnlo} for the final state $X$ in the \PtoB method, where the NNLO corrections take the NLO calculation of $X+j$ production using antenna subtraction as an ingredient:
\begin{align}
  \frac{\rd\sigma_{X}^{\NLO}}{\rd\obs}\; &=\phantom{{}+} 
  \intPhi{n+1}\quad \Bigl( 
    \rd\sigma_X^{R} \bigl( J(\obs_{n+1}) - J(\obs_{n+1\to\Born}) \bigr) \Bigr) 
  + \frac{\rd\sigma_X^{\NLO,\,\incl}}{\rd\obs_\Born} \; , 
  \\[1mm]
  \frac{\rd\sigma_{X}^{\NNLO}}{\rd\obs}\; &=\phantom{{}+} 
  \intPhi{n+2}\quad \Bigl( 
    \rd\sigma_X^{RR}           \bigl( J(\obs_{n+2}) - J(\obs_{n+2\to\Born}) \bigr) 
  - \rd\sigma_{X+j}^{S,\,\NLO} \bigl( J(\obs_{n+1}) - J(\obs_{n+1\to\Born}) \bigr) \Bigr)
  \nonumber\\&\quad
  + \intPhi{n+1}\quad \Bigl(
    \rd\sigma_X^{RV}           \bigl( J(\obs_{n+1}) - J(\obs_{n+1\to\Born}) \bigr) 
  - \rd\sigma_{X+j}^{T,\,\NLO} \bigl( J(\obs_{n+1}) - J(\obs_{n+1\to\Born}) \bigr) \Bigr)
  \nonumber\\[1mm]&\quad
  + \frac{\rd\sigma_X^{\NNLO,\,\incl}}{\rd\obs_\Born} \; .
  \label{eq:p2blow}
\end{align}
The contributions ${\rd \sigma_X^{A}}$ denote the corrections to the Born-level process from single-real radiation $(A=R)$, double-real radiation $(A=RR)$ and real--virtual corrections $(A=RV)$, and mass-factorisation contributions are again implicitly included with the virtual corrections.

\section{Application to jet production in deep inelastic scattering}
\label{sec:dis}

The basic parton-level process in deep inelastic lepton--proton scattering (DIS) is the elastic scattering at large momentum transfer of the lepton and a quark, mediated by a virtual photon, $\PW$- or $\PZ$-boson. 
The outgoing quark forms a jet, of which the momentum can be inferred using momentum conservation using the initial state momenta and the outgoing electron momentum. 
Viewed in the laboratory frame of the lepton--proton system (which can be either of fixed-target or collider type), this jet always has a non-vanishing transverse momentum and a finite rapidity. 
The inclusive single-jet cross section (at fixed kinematics of the jet) at Born level is therefore identical to the inclusive structure function (at fixed kinematics of the lepton). 

Jet production in DIS has been measured~\cite{newman} in the laboratory frame~\cite{e665lab,h1lab,zeuslab,zeus_lab_rate,zeus_lab_diff} and in the Breit frame~\cite{h1breit,zeusbreit}.
In the latter, the virtual photon and the proton collide head-on and the basic lepton--quark scattering process always yields a quark at zero transverse momentum.
Consequently, the first non-trivial jet production process in the Breit frame is two-jet production. 
Owing to the higher final state multiplicity, this process has a higher sensitivity to QCD dynamics; the vast majority of jet production studies at HERA were therefore performed in the Breit frame~\cite{h1breit,zeusbreit}. 

Next-to-leading order QCD corrections to jet production in DIS have been available for a long time for single-jet production in the laboratory frame~\cite{nlolab} as well as for two-jet~\cite{nlo2j} and three-jet production~\cite{nlo3j} in arbitrary frames. 
Very recently, the calculation of NNLO QCD corrections to two-jet production in DIS~\cite{disprl} was completed. 
This calculation uses the antenna subtraction method~\cite{ourant,currie} and is implemented in a flexible parton-level event generator (\NNLOJET), which allows to compute any infrared-safe observable derived from the process under consideration. 
While~\cite{disprl} and its subsequent application to a precision determination of the strong coupling constant~\cite{h1as} discuss only jet production observables in the Breit frame, the very same \NNLOJET implementation can be applied to compute two-jet production in DIS in the laboratory frame. 

The Born level kinematics of single-jet production in DIS in this frame can be reconstructed entirely from the lepton kinematics:
denoting the incoming and outgoing lepton momenta by $p_a$ and $p_1$ (with $q=p_a-p_1$), and the incoming proton momentum by $P$, the incoming and outgoing quark momenta are determined by $p_b=xP$ and $p_2=xP-q$ with $x=-q^2/(2q\cdot P)$. 
The Born-level value of any observable $\obs_\Born \equiv \obs_2$ derived from single-jet production in DIS is therefore expressed in terms of the momenta $(p_a,p_1,P)$. 
As a result, we can define a mapping which, starting from any  momentum set of final-state multiplicity $m > 2$, maps onto to this Born kinematics: $\{p_i\}_m \to \{p_{i,\Born}\}_2$. 
It is then given by
\begin{align}
  p_{1,\Born} &= p_1 \; , &
  p_{2,\Born} &= xP-q \; .
  \label{eq:projection}
\end{align} 
With this mapping at hand and the identification $X=1j$ and $X+j=2j$, Eqs.~\eqref{eq:p2bsubnum}--\eqref{eq:p2blow} provide the \NNNLO corrections to single-jet production in DIS in the laboratory frame, based on the existing NNLO calculation of two-jet production processes~\cite{disprl} and the inclusive \NNNLO DIS structure function~\cite{mvv}. 
It should be noted that this reasoning exploits the specific constraints on the Born-level kinematics in the laboratory frame, and that the \PtoB method can not be applied in the Breit frame. 
\begin{figure}[htbp]
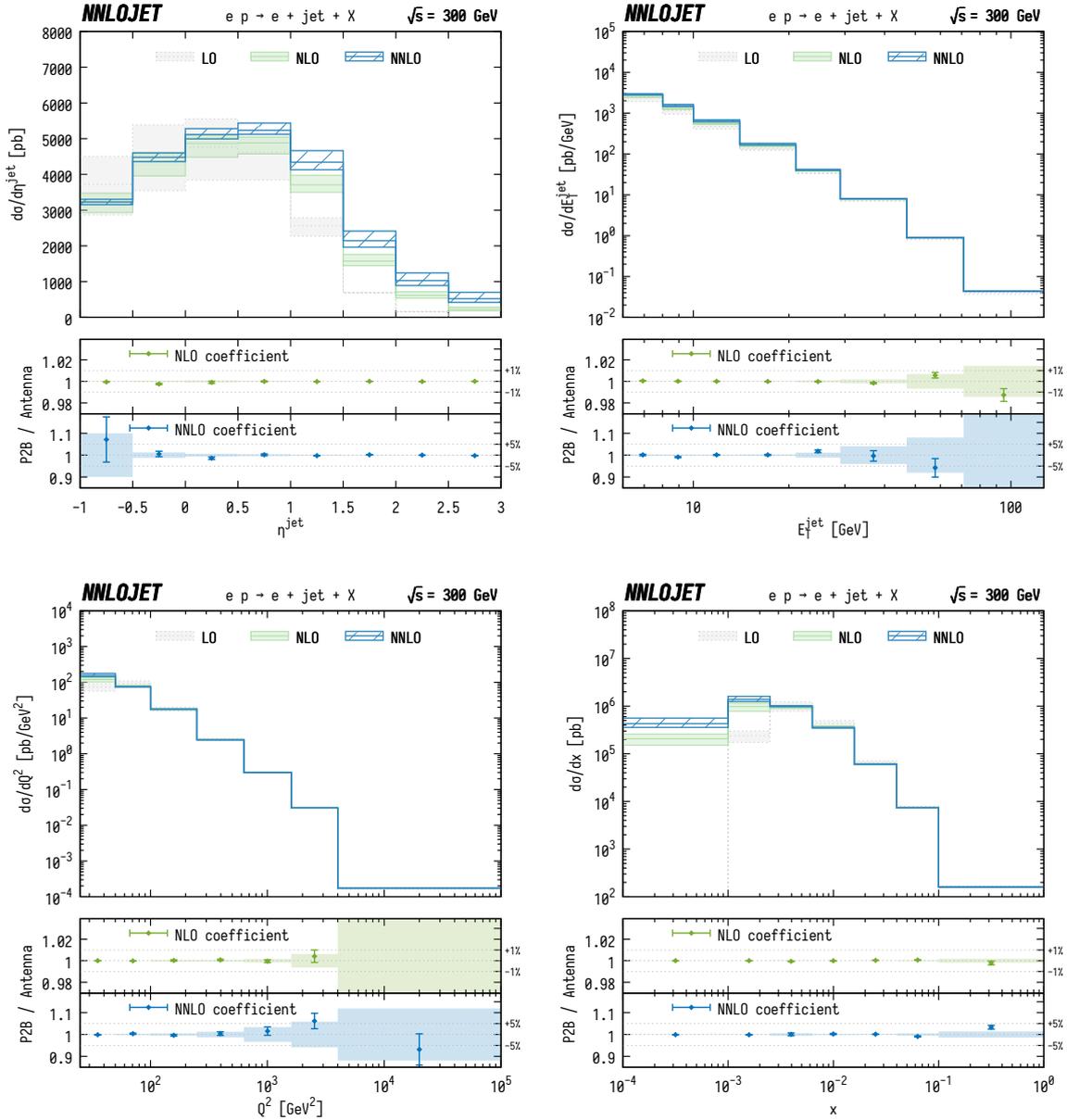

  \includegraphics[width=.49\linewidth]{{{figures/ZEUS_global.validation.etajinc}}} 
  \hfill
  \includegraphics[width=.49\linewidth]{{{figures/ZEUS_global.validation.etjinc}}}
  \\[1em]
  \includegraphics[width=.49\linewidth]{{{figures/ZEUS_global.validation.q2}}}
  \hfill
  \includegraphics[width=.49\linewidth]{{{figures/ZEUS_global.validation.x}}}
  \caption{%
  Comparison of results up to NNLO obtained with the \PtoB and the antenna subtraction methods. 
  The upper frames in each plot show the absolute predictions obtained using the \PtoB method. 
  The lower frames display the ratio of the predictions obtained in the \PtoB and antenna subtraction computations for the NLO-only and NNLO-only contributions to the differential cross sections, with error bars and filled areas representing numerical integration errors of the \PtoB and antenna-subtraction results, respectively.
  The cuts and jet definition are as in Eqs.~\eqref{eq:diff:cuts} and \eqref{eq:diff:jets}.
  \label{fig:P2BAnt}}
\end{figure}

In order to validate our implementation of the \PtoB method up to NNLO, we have performed an independent calculation of single jet production in the laboratory frame based entirely on the antenna subtraction method. 
In Fig.~\ref{fig:P2BAnt}, we compare results from both methods for the inclusive 
jet pseudorapidity $\eta^{\:\!\jet}$, the inclusive jet transverse energy $E_\rT^\jet$, the momentum transfer $Q^2 = -q^2$, and the Bjorken scaling variable $x$.
The setup of the calculation follows the ZEUS measurement~\cite{zeus_lab_diff} described in Sect.~\ref{sec:results:diff} below.
The bottom panels in each plot display the ratios between the \PtoB and antenna subtraction method for the NLO and NNLO \emph{coefficients} and the central choice $\muR=\muF=Q$ of the renormalization and mass-factorization scales.
The error bars correspond to the numerical integration error of the \PtoB prediction, while the filled area shows the corresponding error of the antenna-subtraction results.
The dashed lines are for illustrative purposes and show $\pm1\%$ for NLO-only and $\pm5\%$ for NNLO-only.

We observe agreement between the two methods for the $\ord(\alphas)$ NLO coefficient, which is well below $1\%$ across almost the entire kinematic range and fully consistent within the respective Monte Carlo errors.
For the $\ord(\alphas^2)$ NNLO coefficient, the statistical uncertainties are typically below $1\%$ in regions which contribute the bulk of the cross section; they can increase to a few percent in the tails of the distributions. 
Again, we observe excellent agreement between the two independent calculations. 
With respect to the full NNLO prediction, the agreement between the two methods was validated at the sub-permille level.
A similar level of agreement is also observed for the other scale settings of the seven-point scale variation described below.

\section{\texorpdfstring{\NNNLO}{N3LO} results and comparison to HERA data}
\label{sec:results}

The first observation of jet production in DIS was made by the fixed-target E665 experiment~\cite{e665lab}. 
Shortly thereafter, the experiments H1 and ZEUS at the electron--proton collider HERA embarked on a large program of 
jet production studies in deep inelastic scattering, with measurements both in the laboratory frame~\cite{h1lab,zeuslab,zeus_lab_rate,zeus_lab_diff} and in the Breit frame~\cite{h1breit,zeusbreit}. 
Early measurements established the jet production process and determined total jet rates as function of the 
resolution parameter~\cite{h1lab,zeuslab,zeus_lab_rate}. Subsequent studies on a larger dataset led to the 
determination of differential distributions of jets~\cite{zeus_lab_diff,h1breit,zeusbreit} in the kinematical variables. 
The resulting HERA legacy dataset provides important constraints on QCD dynamics and the parton distributions of the proton. 

In the following, we 
will compare the \NNNLO results to single-jet measurements in the laboratory frame that were 
performed by the ZEUS collaboration for differential distributions~\cite{zeus_lab_diff} and jet rates~\cite{zeus_lab_rate}.
The theory uncertainties on the predictions 
are obtained from a seven-point scale variation of renormalisation and factorisation scales around a central value of $\muf^2=\mur^2=Q^2$, independently varying $\muR$ and $\muF$ up and down by factors $[1/2\,,\,2]$ and discarding the two combinations with the largest separation of both scales.
Parton distribution functions fitted at \NNNLO accuracy are not yet available; the results presented below are obtained using the NNLO NNPDF3.1 set~\cite{nnpdf} with $\alphas(M_\PZ)=0.118$. The same set is  also used to evaluate the
  LO and NLO expressions. 

\subsection{Differential distributions}
\label{sec:results:diff}

The ZEUS measurement~\cite{zeus_lab_diff} of differential distributions for jet production in the laboratory frame is based 
on data that were taken with a proton beam energy $E_\Pp=820~\GeV$ and an electron beam 
energy $E_\Pe=27.5~\GeV$. 
We compare our results to the measurement performed in the `global' region\footnote{The other two regions that are 
presented in the 
ZEUS study are subsets of the `global' region, obtained by applying additional cuts to effectively 
remove the Born-level 
one-jet production process. We do not consider these here.} defined by 
ZEUS through the fiducial cuts
\begin{align}
  Q^2 &> 25~\GeV^2 , &
  y &= \frac{Q^2}{xs} > 0.04 , &
  E_\Pe' &> 10~\GeV ,
  \label{eq:diff:cuts}
\end{align}
where $E_\Pe'$ denotes the energy of the outgoing electron. 
Jets are reconstructed using the $k_\rT$ clustering algorithm~\cite{ktclus}
 in the longitudinally invariant mode ($E_\rT$-weighted recombination scheme)~\cite{ktlong} and are required to satisfy
\begin{align}
  E_\rT^\jet &> 6~\GeV , &
  -1 &< \eta^{\:\!\jet} < 3 \; .
  \label{eq:diff:jets}
\end{align}

\begin{figure}[htbp]
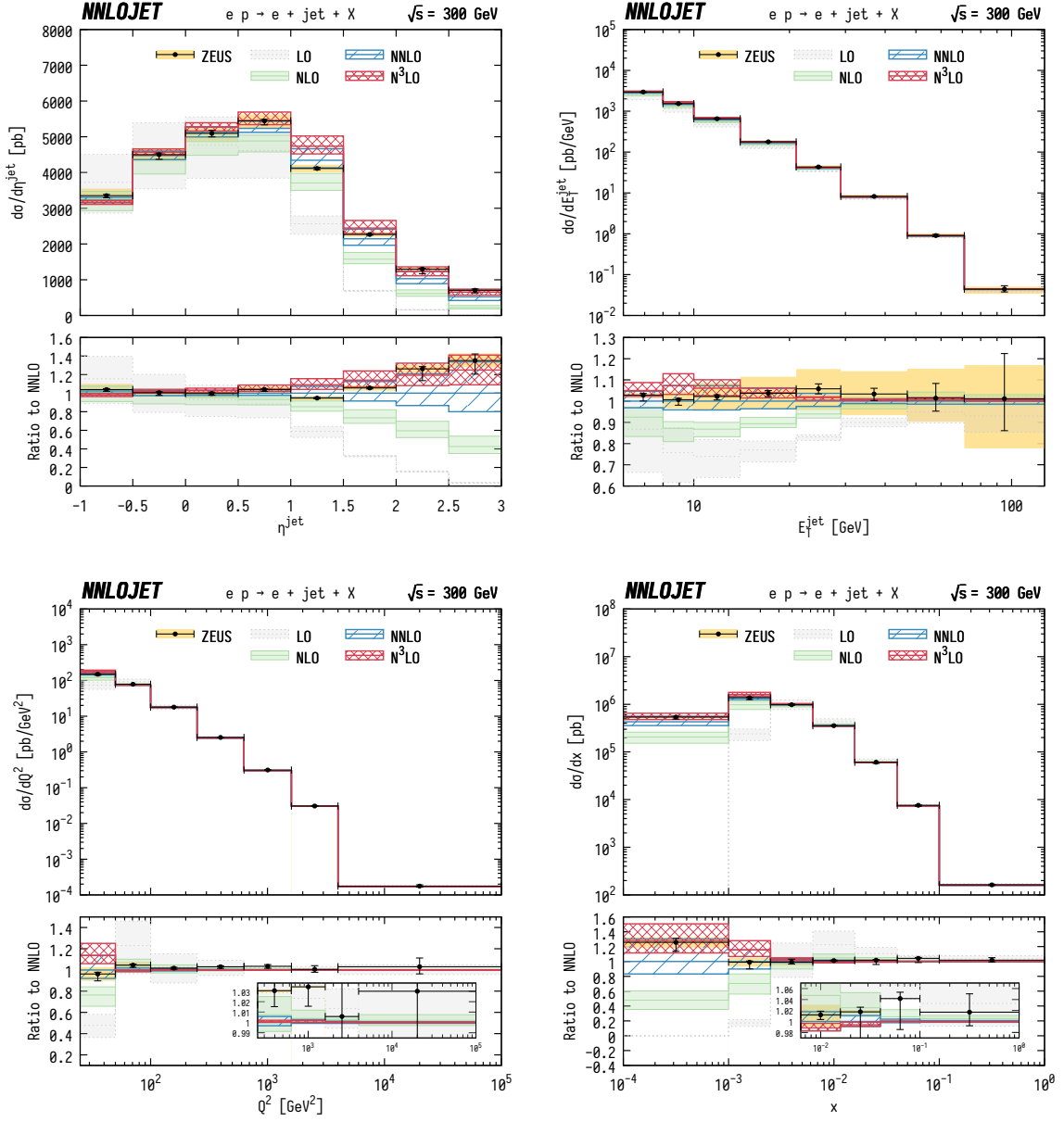

  \includegraphics[width=.49\linewidth]{{{figures/ZEUS_global.etajinc}}} 
  \hfill
  \includegraphics[width=.49\linewidth]{{{figures/ZEUS_global.etjinc}}}
  \\[1em]
  \includegraphics[width=.49\linewidth]{{{figures/ZEUS_global.q2}}}
  \hfill
  \includegraphics[width=.49\linewidth]{{{figures/ZEUS_global.x}}}
  \caption{%
  Kinematical distributions in single inclusive jet production in deep inelastic scattering up to \NNNLO in QCD, compared to ZEUS measurements~\protect\cite{zeus_lab_diff}.
  The error bars on the data represent the statistical and systematic uncertainties added in quadrature; the uncertainty in the absolute energy scale of the jets is shown separately as a shaded yellow band.
  \label{fig:DIS_diff}}
\end{figure}

Figure~\ref{fig:DIS_diff} compares the cross sections calculated at LO, NLO, NNLO and \NNNLO to the experimental measurements~\cite{zeus_lab_diff} for the inclusive jet pseudorapidity $\eta^{\:\!\jet}$, the inclusive jet transverse energy $E_\rT^\jet$, the momentum transfer $Q^2$, and the Bjorken scaling variable~$x$.
We observe that, for the first time, the scale-uncertainty bands overlap across the full kinematic range, when going from NNLO to \NNNLO.
The inclusion of the \NNNLO corrections further reduces the scale uncertainties, by typically a factor of two or more.

The low-$x$ region as well as the first $Q^2$ bins are kinematically suppressed at LO, hence the perturbative accuracy is effectively reduced.
As a consequence, we observe larger higher-order corrections with residual \NNNLO scale uncertainties at the level of 
about $\pm10\%$ in this region.
Large perturbative corrections and correspondingly 
large uncertainties are also observed in the forward region, $\eta^{\:\!\jet} \gtrsim 1$. In this region, the LO process 
is again suppressed kinematically, and most of the jet activity here arises from final states containing several jets. 
Forward jet production in DIS has been studied extensively~\cite{newman} 
with a view on establishing large logarithmic corrections 
arising from the high-energy limit~\cite{smallx} that potentially require resummation.  
We observe that in the forward region these corrections are built up by including consecutive 
perturbative orders. The overlap of the scale uncertainty bands at NNLO and \NNNLO 
indicates a stabilization of the expansion at the present order. 

Overall, the \NNNLO QCD predictions provide an 
excellent description of the ZEUS data. 
An improvement over the NNLO description 
is observed in particular  in those kinematical regions where higher-order corrections are large
(low $x$, low $Q^2$, forward region: $\eta^{\:\!\jet} \gtrsim 1$): 
Here the \NNNLO corrections induce changes to the shape 
which bring the central predictions in line with the measurements. 
The shape of the $E_\rT^\jet$ distribution, especially in the region $8$--$20~\GeV$, is also affected by higher-order corrections; however, its experimental accuracy is systematically limited by relatively large jet-energy-scale uncertainties.

\subsection{Jet rates}
\label{sec:results:rates}

Earlier ZEUS measurements~\cite{zeuslab,zeus_lab_rate}, based on a smaller dataset
taken with $E_\Pp=820~\GeV$  and $E_\Pe=26.7~\GeV$, determined the 
jet production rates, i.e., the fraction of events with a certain jet multiplicity. These studies applied 
the JADE clustering algorithm~\cite{jade} with the four-momentum recombination scheme.
The jet rates were measured as a function of the JADE clustering parameter $y_\cut$ in the fiducial region defined 
by~\cite{zeus_lab_rate}:
\begin{align}
  160  &< Q^2 < 1280~\GeV^2 , &
  0.04 &< y < 0.95 \: , &
  0.01 &< x < 0.1 \: .
 \label{eq:rates:cuts}
\end{align}  
In this particular measurement, the two-jet rate is defined as
\begin{equation}
  R^{\:\!\ZEUS}_{(2+1)} = \frac{ N_{2+1} }{ N_{2+1} + N_{1+1} } \;  ,
\end{equation}
where $N_{1+1}$ and $N_{2+1}$ are the number of recorded $(1\!+\!1)$- and $(2\!+\!1)$-jet events, with the extra ``+1'' denoting the proton remnant forming the beam jet. 
The normalisation is chosen such that 
\begin{equation}
  R^{\:\!\ZEUS}_{(1+1)} \equiv 1-R^{\:\!\ZEUS}_{(2+1)} \; ,
  \label{eq:JRNORM}
\end{equation}
which is different from the usual convention to normalise with respect to the total hadronic cross section.

\begin{figure}[htbp]
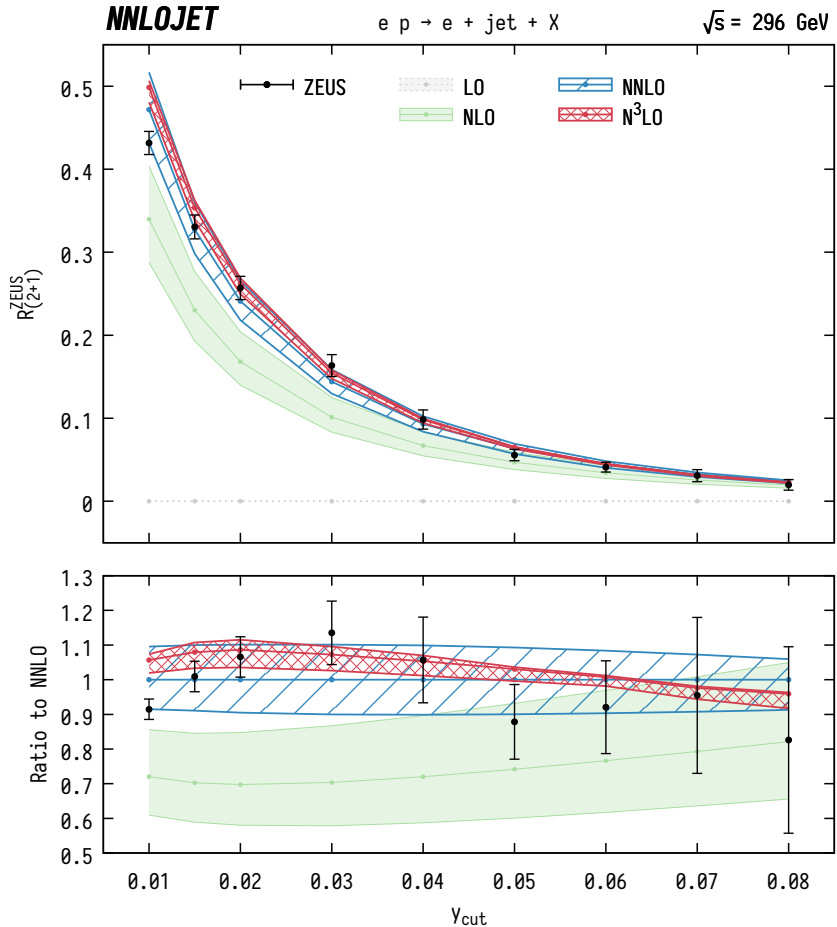

  \centering
  \includegraphics[width=0.75\linewidth]{{{figures/ZEUS.rates_2}}}
  \vspace*{-2mm}
  \caption{%
  Comparison of ZEUS data to theoretical predictions up to \NNNLO in QCD for the two-jet rate, normalised according to Eq.~\eqref{eq:JRNORM}. 
  Data are corrected to give parton level results and were extracted from Fig.~9 of Ref.~\cite{zeus_lab_rate}; the error bars show the statistical uncertainties.
  \label{fig:DIS_ZEUS}}
\end{figure}

In Fig.~\ref{fig:DIS_ZEUS} we present the results for the 2-jet rate $R^{\:\!\ZEUS}_{(2+1)}$ up to \NNNLO and compare them to the measurement in Ref.~\cite{zeus_lab_rate}. 
We refrain from showing the one-jet rate separately, as it is trivially related to $R^\ZEUS_{(2+1)}$ via Eq.~\eqref{eq:JRNORM}.
It should be noted that the errors on the data~\cite{zeus_lab_rate}
correspond only to the statistical uncertainties. 
Systematic uncertainties from jet acceptance corrections (which amount to up to 20\%, and whose uncertainty 
is not quantified) are not provided on a 
bin-by-bin basis in~\cite{zeus_lab_rate}. 

It can be seen that the \NNNLO corrections result in a substantial reduction of scale uncertainties.
As the value of $y_\cut$ is lowered, less of the final-state radiation is clustered, thereby resolving more 
jet structure. 
As a result, the fractions of two-jet events increases towards lower values of $y_\cut$.
The low-$y_\cut$ region is also where the largest scale dependence is observed. 
At \NNNLO the scale band starts to overlap in the low-$y_\cut$ region with that of the previous order for the first time.
This is accompanied with a substantial reduction of scale uncertainties, and the $y_\cut$-dependence of the 
  data is  well-described by the predictions at NNLO and \NNNLO. 

\begin{figure}[htpb]
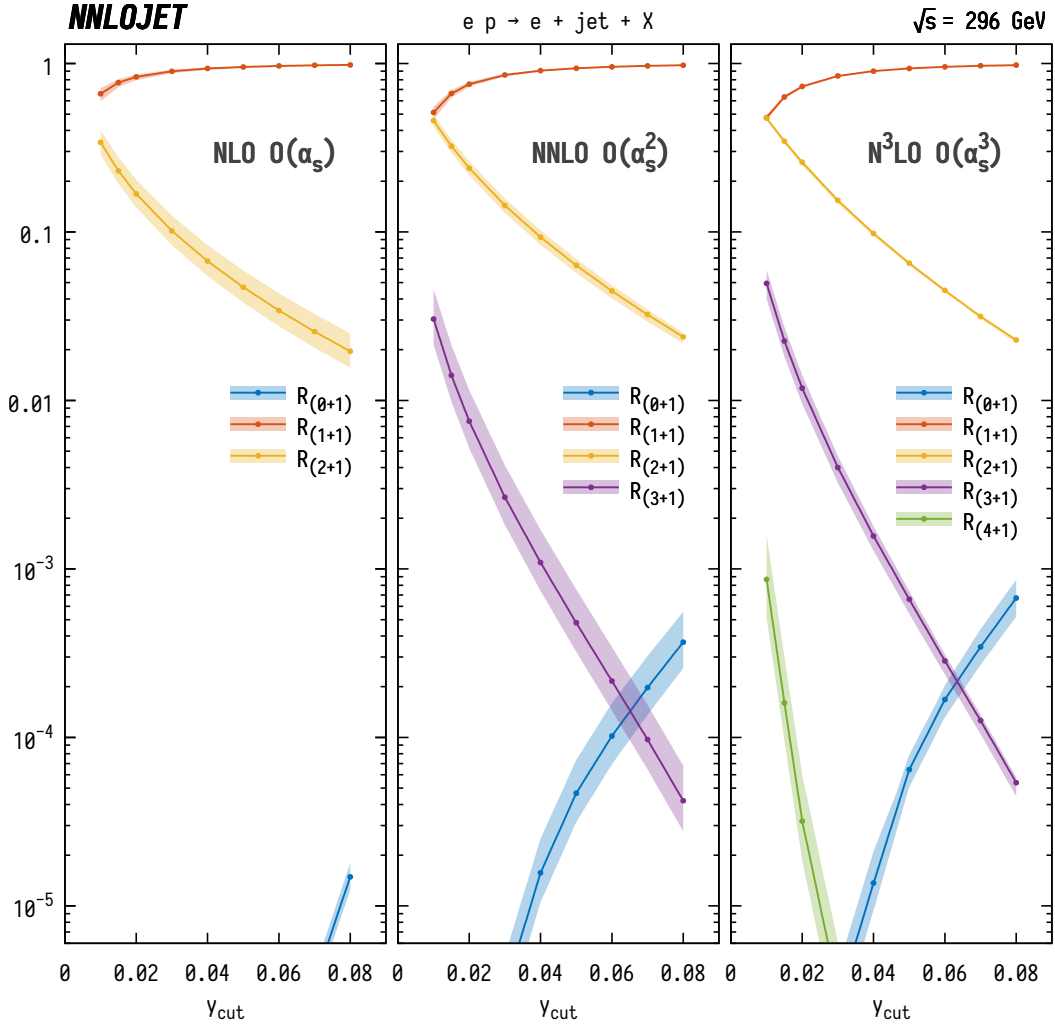

  \centering
  \vspace*{-1mm}
  \includegraphics[width=0.95\linewidth]{{{figures/DIS.rates}}}
  \vspace*{-7mm}
  \caption{%
  Predictions up to \NNNLO in QCD for zero-, one-, two-, three-, and four-jet rates in the JADE algorithm for 
  the kinematical cuts of Eq.~\eqref{eq:rates:cuts}.
  \label{fig:DIS_rates}}
\end{figure}

Using the same 
kinematical cuts, we have also calculated the jet rates normalising to the total hadronic cross section, i.e.\
not applying the unusual normalisation in Eq.~\eqref{eq:JRNORM}. In this setting,
 the zero-, three-, and four-jet rates are also well-defined and can  be evaluated, as shown in Fig.~\ref{fig:DIS_rates}. 
Similar features as in Fig.~\ref{fig:DIS_ZEUS} are observed
concerning  the reduction of the scale 
uncertainty and the $y_\cut$-dependence. Moreover, it can be seen that the 
 fractions of one- and two-jet events dominate over events with higher jet multiplicities by at 
 least an order of magnitude.

\pagebreak

\section{Conclusions}
\label{sec:conc}

In this paper, we demonstrated how the \PtoB method for infrared subtractions can be applied to 
compute fully differential third-order (\NNNLO) QCD corrections to observables with sufficiently simple Born-level 
kinematics. The implementation of the method for a given process $X$ requires the knowledge of the 
fully inclusive \NNNLO cross section for the production of $X$, and a fully differential NNLO calculation 
for $X+$jet final states. 

As a first application, we have computed the \NNNLO corrections to jet production in deep inelastic 
lepton--proton scattering. Predictions at this order lead to very small residual scale uncertainties (often at
 sub per-cent level 
in the bulk of the phase space) and account properly for enhancements of distributions from multiple radiation 
near 
boundaries of the phase space. The \NNNLO results provide an excellent description of data 
on distributions and rates in jet production, obtained by the
ZEUS experiment~\cite{zeus_lab_rate,zeus_lab_diff}, also in those kinematical regions where the NNLO 
predictions were insufficient to describe their behaviour. 

The \PtoB method could be used in the near future to 
evaluate more processes at \NNNLO accuracy in particular in proton--proton collisions.

\acknowledgments

The authors are grateful to Mark Sutton for bringing the ZEUS 
measurements~\cite{zeuslab,zeus_lab_rate,zeus_lab_diff} to our attention and his help on 
questions regarding the details of the analyses.
We further thank Xuan Chen, Juan Cruz-Martinez, Rhorry Gauld, Aude Gehrmann--De Ridder, Marius Höfer, Imre Majer, Tom Morgan, Joao Pires, Duncan Walker and James Whitehead for useful discussions and their many contributions to the \NNLOJET code.
We acknowledge the computing resources provided to us by the Swiss National Supercomputing Centre (CSCS) under the project ID p501b.
This research was supported in part by the UK Science and Technology 
Facilities Council, by the Swiss National Science Foundation (SNF) under contracts 200020-175595 
and CRSII2-160814, and by the Research Executive Agency (REA) of the European Union through the ERC Advanced Grant MC@NNLO (340983).



\end{document}